\documentclass[aps,floatfix,twocolumn,amsmath]{revtex4}
\usepackage{graphicx}
\newcommand{\be}{\begin{equation}}
\newcommand{\ee}{\end{equation}}
\newcommand{\bea}{\begin{eqnarray}}
\newcommand{\eea}{\end{eqnarray}}
\newcommand{\no}{\noindent}

\begin{document}


\title{Stabilization of Extra Dimensions \\ and \\ The Dimensionality of the Observed Space}


\author{Tongu\c{c} Rador}
\email[]{tonguc.rador@boun.edu.tr}

\affiliation{Bo\~{g}azi\c{c}i University Department of Physics \\ 34342 Bebek, \.{I}stanbul, Turkey}


\date{\today}

\begin{abstract}
We present a simple model for the late time stabilization of extra dimensions. The basic idea is that brane solutions wrapped around extra dimensions, which is allowed by string theory, will resist expansion due to their winding mode. The momentum modes in principle work in the opposite way. It is this interplay that leads to dynamical stabilization. We use the idea of
democratic wrapping \cite{art5}-\cite{art6}, where in a given decimation of extra dimensions, all possible winding
cases are considered. To simplify the study further we assumed a symmetric decimation in which
the total number of extra dimensions is taken to be $Np$ where N can be called the order of the decimation. We also assumed that extra dimensions all have the topology of tori. We show that with these rather conservative assumptions, there exists solutions to the field equations in which the extra dimensions are stabilized and that the conditions do not depend on $p$. This fact means that there exists at least one solution to the asymmetric decimation case. If we denote the number of observed space dimensions (excluding time) by $m$, the condition for
stabilization is $m\geq 3$ for pure Einstein gravity and $m\leq 3$ for dilaton gravity massaged by a string theory parameter, namely the dilaton coupling to branes.

\end{abstract}


\maketitle

\section{Introduction}

String theory requires for its consistency extra dimensions that are bound to be very small {\em{compared}} to the size of observed dimensions. It is therefor of considerable importance to 
look for cosmological models in which this can be realized or at least not violently denied. String theory also allows compact objects (branes) which could be wrapped around compact extra dimensions. These branes' winding modes will in general resist expansion in the same way as a
rubber band would resist expansion of a balloon around which it is wrapped. The  momentum (vibration) modes on the other hand would tend to enlarge the size of the brane. These two forces might in principle yield stabilization of the extra dimensions realized in a dynamical way. This idea is reminiscent of the Brandenberger-Vafa mechanism presented in \cite{art8}. 
In this letter we enlarge the mass of knowledge on a model which was in development during the past couple of  years \cite{art1}-\cite{art7}.  Interested reader should also check the literature on
brane gas cosmology \cite{art9}-\cite{art27}. For further references one could check a recent review on the topic of brane gas cosmology \cite{art28}. These approaches are in essence different from the brane-world scenarios (see for example \cite{artbw1}  for a review on the topic within the context of cosmology) which also make use of extra dimensions and try to reformulate the hierarchy problem in particle physics and other phenonema.

An interesting puzzle which could be addressed in any theory involving extra dimensions is the dimensionality of
the observed space. This would mean to have an understanding of the question ``Why the number of large dimensions is three?''. In this paper we address the possibility that the requirement of the stability (that is
none or very small present cosmological evolution) of extra dimensions in general and plus that of the dilaton in the context of string theory might give clues to the question. These stabilities are implied by positive observations on the none or very small cosmological evolutions of coupling constants which are supposed to change if the extra dimensions or the dilaton were evolving. The experimental bounds are rather stringent so it makes sense
to look for theories which incorporate absolute stability of extra dimensions and of the dilaton at least at the classical level. To this end we study the brane model we present in the context of pure Einstein gravity and then in the context of dilaton gravity. The bounds on the dimensionality $m$ of observed space are different in both theories (we have $m\geq 3$ for pure Einstein gravity and $m\leq 3$ for dilaton gravity with the dilaton coupling to branes fixed by string theory) and agree only when $m=3$ which also coincides with a stabilized dilaton. That is assuming the extra dimensions are stabilized one can think of Einstein gravity emerging from dilaton gravity  only when $m=3$. On the other hand taking the dilaton to be a constant at the outset is not necessarily compatible with the stability of extra dimensions for a general choice of $m$. This result is interesting in its own and we believe it is a rather solid result for explaining why the number of
observed dimensions is three within the context of extra dimensions and string theory. Unfortunately this mechanism is not a dynamical one as the one advocated in \cite{artbs1} where the authors claim to explain the fact that $m=3$ within the dynamics of string theory \footnote{After the present manuscript appeared Randall and Karch  \cite{artbs2} came up with another variant \cite{artbs1} of  why we live in three dimensions. The present manuscript is not along the same line of reasonning, we simply expose that $m=3$ is the only possibility if one also requires the stability of extra dimensions and the dilaton. We simply assume that ``some'' number of dimensions are somehow singled out at the outset but keep $m$ as a parameter to be fixed later on.}.

Therefor we arrive at the concluding result: since experimentally the stability of coupling constants require the stability of extra dimensions and the dilaton we have $m=3$.

To make things simpler we assume that the extra dimensions (however they are partitioned as product spaces) are tori and hence flat and compact. In view of the lack of a general principle
which would mandate a given wrapping pattern we use democratic winding introduced in \cite{art5} and \cite{art6}. To make
things clearer let us proceed with an example: say extra dimensions are partitioned into three tori of dimensions $p$, $q$ and $r$. The democratic winding scheme require we allow for
all possible windings and hence intersections. Namely the winding pattern will be as follows

\[
(p)qr\oplus p(q)r \oplus pq(r)\oplus (pq)r\oplus p(qr)\oplus q(rp)\oplus (pqr) \;\;.
\]
Here a parenthesis means that there is a brane of dimensionality equal to the sum of dimensions
around which it wraps. For instance $(p)qr$ stands for a $p$-dimensional brane wrapping only around the first partition, $q(rp)$ means there is a $p+r$ dimensional brane wrapping around the
first and last partition and $(pqr)$ is a $p+q+r$ dimensional brane covering all extra dimensions.
For a general decimation pattern the model is complicated. However if we can show that there is a solution for a symmetric decimation in which the dimensionality of the partitionings
are all the same (say $p$) and the total number of extra dimenions is $Np$ and that this solution is $p$ independent it will in general mean that there is at least one solution to the asymmetric partitioning case. 
The ideas of winding democracy and symmetric partitioning were introduced in \cite{art5} for pure Einstein gravity and in \cite{art6} for dilaton gravity. It was shown that the stabilization conditions are $p$ independent in both cases. However in \cite{art5} brane momentum modes were not considered and there remained an $N$ dependence on the stabilization conditions whereas in \cite{art6} it was shown that the results really does not depend on $N$ if one also considers momentum modes. The main idea of this paper is first to add momentum modes to pure Einstien gravity case. And coherently study the two models in such a way to finally contrast the conditions imposed on the dimensionality of observed space by the requirement of stabilization of extra dimensions.

\section{General Formalism}

Since we are interested in a cosmological model we take our metric to be

\be
ds^{2}=-dt^{2}+e^{2 B(t)} dx^{2} + \sum_{i} e^{2 C_{i}(t)} dy_{i}^{2}\;\;.
\ee

Here $B$ stands for the scale factor of the observed space with dimensionality $m$. The $C_{i}$ are the scale factors of extra dimensions. There are $N$ such factors each corresponding to $p$-dimensional tori. The total dimensionality of space-time is $d=m+1+Np$. Because of the symmetric decimation pattern we can take all $C_{i}$ to behave the same way.

We will also assume that the branes are distributed as a continuous gas with respect to the 
directions they are not wrapping. This makes it possible to use a dust-like energy-momentum
tensors. That is for any energy-momentum source $\lambda$ we assume the following form for the energy density
which is found by the conservation requirement of the energy-momentum tensor for that particular source\footnote{So the total energy momentum tensor will be a sum of such separately conserved contributions},

\be
\rho_{\lambda}=\rho^{o}_{\lambda} \exp\left[-(1+\omega_{B}^{\lambda})mB-\sum_{i} (1+\omega_{C_{i}}^{\lambda})p C_{i}\right]
\ee

Here $\omega$'s are the pressure coefficients. It was shown in \cite{art1}-\cite{art4} that if we consider a homogeneous gas of branes the winding mode of a $p$-brane will yield a conserved dust-like energy-momentum tensor with pressure coefficient $-1$ along the winding directions and $0$ for others. For example the energy-densities of the winding modes of a $p$-brane, a $2p$-brane and an $Np$-brane will respectively be

\bea
\rho_{p}&=&\rho^{o}_{p}   e^{-mB}  e^{-(N-1)pC} \nonumber\\
\rho_{2p}&=&\rho^{o}_{2p} e^{-mB}  e^{-(N-2)pC} \nonumber\\
\rho_{Np}&=&\rho^{o}_{Np} e^{-mB}  \nonumber
\eea

For momentum modes the pressure coefficient of a $p$-brane can be taken to be $1/p$ along the
winding directions and vanishing for the rest \cite{art2}-\cite{art4}. So the momentum mode energy-densities for the above list will be

\bea
\tilde\rho_{p}&=&\tilde\rho^{o}_{p}   e^{-mB}  e^{-C-NpC} \nonumber\\
\tilde\rho_{2p}&=&\tilde\rho^{o}_{2p} e^{-mB} e^{-C-NpC} \nonumber\\
\tilde\rho_{Np}&=&\tilde\rho^{o}_{Np} e^{-mB} e^{-C-NpC} \nonumber
\eea

This very simple behaviour of the momentum modes will be a crucial ingredient in proving stabilization. We have also adopted a convention for the initial values of the energy densities: $\rho^{o}$'s will correspond to winding modes and $\tilde\rho^{o}$'s will correspond to momentum modes. 

\subsection{Pure Einstein Gravity}

The field equations for pure Einstein gravity are

\be
R_{\mu\nu}-\frac{1}{2}R g_{\mu\nu}=\kappa^{2} T_{\mu\nu} \;.
\ee

\no With our assumptions the equations of motion for the scale factors can be cast as follows (we set $\kappa^{2}=1$),

\begin{subequations}
\bea
\dot{A}^2 &=& m \dot{B}^2 +\sum_{i} p_{i} \dot{C_{i}}^2 + 2\rho \;, \label{aninki1}\\
\ddot{B}+\dot{A} \dot{B} &=& T_{\hat{b} \hat{b}}-\frac{1}{d-2} T \;,\\
\ddot{C_{i}}+\dot{A} \dot{C_{i}} &=& T_{\hat{c}_{i}\hat{c}_{i}} - \frac{1}{d-2} T\;, \\
A &\equiv& m B +\sum_{i} p_{i} C_{i} \label{aninki2}\;. 
\eea
\end{subequations}

\no The hatted indices refer the the orthonormal co-ordinates. Also $\rho$ represents the total energy density and $T_{\hat{\mu}\hat{\nu}}$ are the components of the total energy-momentum tensor while $T$ is its trace.

Stabilization of extra dimensions will imply

\be
T_{\hat{c}_{i}\hat{c}_{i}} - \frac{1}{d-2} T=0\;\;.
\ee

Considering all the energy-momentum tensors in a democratic winding scheme will yield the following after straightforward algebra

\be\label{poly1}
e^{-mB-NpC}\left[\frac{1}{p}\alpha X^{-1} + \frac{1}{d-2}\sum_{k=1}^{N}\beta_{k}\zeta_{k} X^{kp}\right]=0\;\;.
\ee

with $X\equiv e^{C}$, the scale factor of the extra dimensions, and

\begin{subequations}\label{betas}
\bea
\alpha&=&\sum_{i=1}^{N-1}\tilde\rho^{o}_{ip}+\tilde\rho^{o}_{Np}/N \\
\beta_{k}&=&\rho^{o}_{kp} \\
\beta_{N}&=&\rho^{o}_{Np}/N \\
\zeta_{k}&=&N-k(m-1)
\eea
\end{subequations}

\noindent The difference in the definition of $\beta_{N}$ is due to the fact that there is only
one brane wrapping over all extra dimensions.

To show that there is stabilization we have to find positive solutions to the polynomial in (\ref{poly1}). In order to study this we can use Descartes' sign rule which states that the
positive roots of a polynomial is either equal to the number of sign changes $s$ of the coefficients or less than $s$ by a multiple of 2. Since $\alpha$ and $\beta_{k}$ are all positive numbers the sign changes will be ruled by $\zeta_{k}$. But $\zeta_{k}$ are monotonically decreasing by $k$ for a given $m$, so there can only be one sign change in the 
polynomial (\ref{poly1}) and hence only one positive root exists. The worst case therefor is
given by $\zeta_{N}\geq 0$ which would mandate every term to be positive. This means to have
a sign change we need $m\geq 2$, however for $m=2$, $\zeta_{N}$ is zero and $\zeta_{N-1}>0$. 

Consequently the real constraint to have a solution is  

\be 
m\geq 3. 
\ee

This result does not depend on $N$ or on $p$ so there must be at least one solution for stabilization even in the (very difficult to analyze) case for which the decimation of extra dimensions is not symmetric. It can also be shown that with these stabilization conditions the observed space expands with the same power-law (2/m) as presureless dust. This is expected since all the brane energy-momentum tensors are
presureless dust for the observed space.

\subsection{Dilaton Gravity}

We can take the action in the presence of dilaton field $\phi$ coupled to matter to be \cite{art4},

\be
S=\frac{1}{\kappa^{2}}\int dx^{d}\;\sqrt{-g}\;e^{-2\phi}\left[R + 4 (\nabla \phi)^2 + e^{a\phi}\mathcal{L}_{m}\right]\;.
\ee

If the $\mathcal{L}_{m}$ takes the form of a lagrangian yielding a dust-like energy-momentum
tensor the field  equations are \cite{art4}, 

\begin{subequations}
\bea
R_{\mu\nu}+2\nabla_{\mu}\nabla_{\nu}\phi = &e^{a\phi}&\left[T_{\mu\nu}-(\frac{a-2}{2})\rho g_{\mu\nu}\right] \\
R+4 \nabla^{2}\phi -4 (\nabla\phi)^{2} &=& -(a-2)e^{a\phi}\rho\;.
\eea
\end{subequations}

\noindent which in turn will give the following, (we set $\kappa^{2}=1$),

\begin{subequations}
\bea
\ddot{B}&=&-k\dot{B}+e^{a\phi}\left[T_{\hat{b}\hat{b}}-\tau\rho\right]\;,\label{bieq}\\
\ddot{C_{i}}+k\dot{C_{i}}&=&e^{a\phi}\left[T_{\hat{c_{i}}\hat{c_{i}}}-\tau\rho\right]\;,\label{cieq}\\
\ddot{\phi}&=&-k\dot{\phi}+\frac{1}{2}e^{a\phi}\left[T-(d-2)\tau\rho\right]\;,\label{phieq}\\
k^{2}&=&m\dot{B}^{2}+\sum_{i}p_{i}\dot{C_{i}}^{2} + 2 e^{a\phi}\rho\;,\label{refk1} \\
k&\equiv& m\dot{B}+\sum_{i}p_{i}\dot{C_{i}}-2\dot{\phi}\;.\label{refk2}
\eea
\end{subequations}

\noindent here $\tau=(a-2)/2$. The stabilization condition is

\be
e^{a\phi}\left[T_{\hat{c_{i}}\hat{c_{i}}}-\tau\rho\right]=0\;\;.
\ee

Similarly to the previous subsection, after considering all the energy-momentum contributions, this will 
yield the following

\be\label{poly2}
e^{a\phi-mB-NpC}\left[(\frac{1}{p}-\tau N)\alpha X^{-1}-\sum_{k=1}^{N}\beta_{k}\xi_{k}X^{kp}\right]=0
\ee

Here $\alpha$ and $\beta_{k}$ are the same parameters of  the previous subsection, as in (\ref{betas}), and $\xi_{k}=k+\tau N$. The discussion for a solution is very similar to the previous section. There can only be one sign change in the polynomial (\ref{poly2}) due to the linear change in $\xi_k$. It is easy to show that solutions will exist for

\be
-1<\tau<\frac{1}{Np}\Longrightarrow -1<\tau<\frac{1}{d-m-1}
\ee

Again since the constraints do not depend on $N$ or $p$ and this means there should at least be one solution to the stabilization conditions for asymmetric decimations. 

Furthermore in \cite{art6} it has been shown that the observed space's scale factor and the dilaton evolve according to a power-law ansatz,
\begin{subequations}
\bea
B(t)\sim \beta \ln{t}\;,\\
\phi(t)\sim \varphi \ln(t)\;.
\eea
\end{subequations}

\no with,

\begin{subequations}
\label{bphi}
\bea
\beta&=&-\frac{2\tau}{1+(m-1)\tau^{2}}\label{bphib}\;,\\
\varphi&=&\frac{-2+m\beta}{2(1+\tau)}=-\frac{1+\tau (m-1)}{1+(m-1)\tau^2}\label{varphib}\;.
\eea
\end{subequations}

Since we want to use these as late time cosmology solutions we would like to have the scale of the observed space expanding. Also to not enter the strong coupling regime of string theory at late times we would like to have a decreasing (or stable) dilaton solution\footnote{The requirement for a decreasing or stable dilaton is the phenomenologically favored situation. One could in principle look for the increasing dilaton case and this can prove to be of interest in another context. We will not pursue this idea here}. Thus we want $\beta>0$ and $\varphi\leq0$ in the equation above. These further requirements will alter the stabilization conditions in the following way

\be
-\frac{1}{m-1}\leq\tau<0\;\;.
\ee

In string theory $\tau=-1/2$ for $Dp$-branes. This in turn means

\be
m\leq3
\ee

\subsection{Stability}

The pure Einstein and Dilaton gravity cases share a common property for the evolution of the extra dimensions. The equations governing the behaviour of the scale factors of extra dimensions is always in the following form 

\be
\ddot{C}=-f(t,\dot{C})\dot{C}+g(t)\;X^{-Np-1} P(X)
\ee

\no with again $X=e^{C}$. The polynomial $P(X)$ has a single positive root. The general structure of this polynomial is as follows, 

\be\label{pol}
P(X)=1+a_{o}X+a_{1}X^{p+1}+\cdots+a_{N}X^{Np+1}
\ee

\begin{figure}[t]
\includegraphics[scale=0.33]{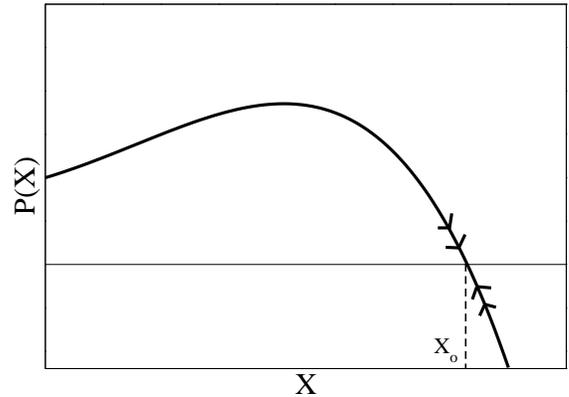}
\caption{{\label{fig:a1}} The generic form of the stabilization polynomial in (\ref{pol}).}
\end{figure}

\no where the constant term comes from the collective momentum modes, the linear term comes from ordinary pressureless matter living in the observed space (and hence has zero pressure coefficients everywhere) and the terms involving powers of $p$ are coming from the winding modes. The condition for stabilization is that after/before the $k$'th term all $a_{k}$'s are negative/positive. Therefor invoking Descartes' rule again we would have unique solutions for the vanishing of the derivatives up to the $k-1$'th order. Thus $P(X)$ increases starting from $X=0$ and starts decreasing after the unique solution to $P'(X)=0$ until it reaches the unique stabilization point $P(X_{o})=0$. 

On the other hand the function $g(t)$ is always positive \footnote{since it is just a positive factor times $e^{-mB}$}. As it stands the equations is a (non-linearized) motion of a particle under the influence of a position dependent force and a velocity dependent friction/driving force $f$. As one could guess if $f<0$ the stabilization might be jeopardized and it does, this fact have been numerically substantiated in \cite{art4}: if $f<0$ there is a singularity in the field equations in finite proper time. If one takes a good look at the field equations in either pure Einstein of Dilaton gravity the sign of $f$ is a constant of motion \footnote{This follows from the defitions of $\dot{A}$ and $k$ in equations (\ref{aninki1}),(\ref{aninki2}) and (\ref{refk1}),(\ref{refk2}) respectively}. We therefor should consider the $f>0$ case.

As for the position dependent force we can look for the potential that gives rise to it

\be{\label{veff}}
-\frac{dV_{eff}}{dX}=X^{-Np-1}P(X)
\ee

\begin{figure}[t]
\includegraphics[scale=0.33]{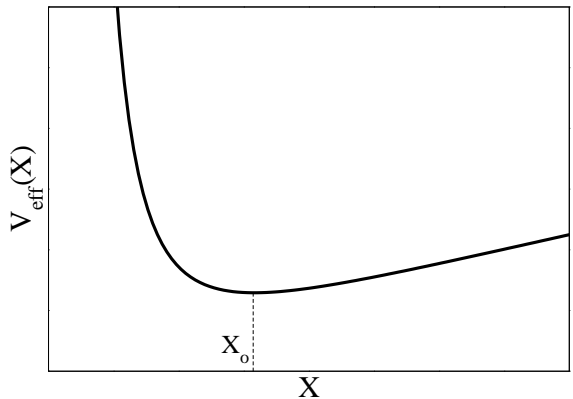}
\caption{{\label{fig:a2}} The generic form of the effective potential defined in (\ref{veff}).}
\end{figure}

The general form of $P(X)$ and $V_{eff}(X)$ are represented in the figures \ref{fig:a1} and \ref{fig:a2}. The potential has a unique minima and therefore the stabilization solutions are truly stable since $f>0$ will just bring in the friction. The effective potential has a very strong repulsive core near the small $X$ region in the form inverse powers of $X$, the strongest one coming from the collective momentum modes. The large $X$ behaviour is dominated by the winding mode of largest brane in the system, in a democratic wrapping scheme this would be a term linear in $X$ and comes from the $Np$-brane that wraps the entire decimation. These results are very plausible, the momentum modes are the ones that resist the contraction the most  and the largest brane's winding mode is the one that resist expansion the most.

We thus have shown that the stabilization solution is a future attractive point given $f(t=0)>0$ and the
internal dimensions will evolve to that point no matter what the initial conditions are \footnote{The stability analysis we have exposed here is only valid in the classical regime. The ultimate analysis should have included the effects of quantum fluctuations which we do not pursue here since it would require an all out use of string theory. We advocate on the
other hand that the classical stability is a strong argument since it actually becomes more and more valid at
later times when string theory becomes less and less important.}. It has also been recently shown by Kaya \cite{art7} that the stabilization point is also dynamically stable against cosmological perturbations of the metric.

\subsection{Conclusion}

The analysis of the previous chapter can be summarized as follows. The late time stabilization of extra dimensions by dynamics of $Dp$-branes require 

\begin{subequations}
\bea
m\geq &3& \rm Pure\;Einstein\;Gravity.\nonumber\\
m\leq &3& \rm Dilaton\;Gravity\;(String\;Theory\;input\;a=1).\nonumber
\eea
\end{subequations}

The only case in which they would both agree is the experimentally observed case, $m=3$. In this case the dilaton is also stabilized and the observed space expands as the ordinary presureless dust solution with power $2/3$. The fact that the dilaton stabilizes means that the Einstein frame and the String frame are the same in the far future. One of the reasons why two different models yield different regimes for stabilization
is that one should not really think that pure Einstein gravity can be obtained from dilaton gravity by setting the dilaton to a constant because the evolution equation for the dilaton is not necessarily satisfied identically for every parameter of the system. However one can think of obtaining pure Einstein gravity from dilaton gravity by setting the dilaton to a constant when $m=3$. 

Since one would like to recover
Einstein gravity for low energies it seems $m=3$ is mandated by stabilization of extra dimensions.

Although we have confined the present study to the symmetric decimation of the extra dimensions (where each one having dimensionality $p$), the fact that the stabilization condition does not depend on $p$ means  there should be at least one solution to the stabilization equations in the asymmetric decimation case. So the mechanism is generic. 

It is also rather interesting that in the dilaton case we have found $m\leq 3$ without requiring that in the early universe $p$-branes with $p>2$ annihilated as one would argue in the case of Brandenberger-Vafa mechanism. If one would like to apply  this constraint of the Brandenberger-Vafa mechanism to the model of this work no part of a decimation can have $p>2$, and there
is simply no solution for stabilization in this case.

One important point to mention is the possibility to allow for internal curvature for extra dimensions. An in depth study is not within the scope of this manuscript, however a simply analysis shows that the internal curvature frustrates the mechanism we presented. The reason being the curvature will bring a factor of $2ke^{-2C}$ to the stabilization polynomials and since this term has no overall $e^{-mB}$ factor, stabilization can not be achieved as we laid out. On the other hand this can also be interpreted as a hint that extra dimensions
have no curvature. 

One immediate extension of the work presented here is to add pressureless dust (galaxies, quasars etc.) in the observed dimensions. Since this form of matter will bring a positive term with a factor $e^{-mB}$ to the stabilization polynomial the mechanism we have presented will work out as well at least for Einstein gravity. For dilaton gravity on the other hand one will have to know the coupling $a_{matter}$ of the dilaton to ordinary matter and this
is not known. However if the dilaton stabilization occurs before matter domination nothing will change.

Other important extensions one has to study are the study of the model during the radiation and early inflationary eras. Early inflation tends to grow extra dimensions as well as the observed dimensions unless
one introduces rather contrived and unmotivated cancellation mechanisms and consequently during radiation a shrinking has to occur for extra dimenisions for them to get as small as the experimental bounds. A preliminary analysis shows this to be true but we leave an extensive study  for a future work. This bounce back of the size of extra dimensions can be very useful if used together with the bounds on the change of the fine structure constant coming from primordial nucleosynthesis. There could also be interesting avenues if one includes the present accelaration of the universe.

\section{Appendix}

In this appendix for didactical purposes we would like to expose  the simpler case of $N=1$. That is we assume that the extra dimensions are lumped in a $p$-dimensional torus. The energy densities for the
brane winding and momentum modes will be given as

\begin{subequations}
\begin{eqnarray}
\rho_{p}&=&\rho_{p}^{o} e^{-mB}\\
{\tilde\rho}_{p}&=&{\tilde\rho_{p}}^{o} e^{-mB-(1+p)C} 
\end{eqnarray}
\end{subequations}

\noindent{\bf{Pure Einstein Gravity}}

The equation for the evolution of extra dimensions will be 

\be
\ddot{C}+\dot{A}\dot{C} = -\frac{m-2}{d-2} \rho_{p} + \frac{1}{p} {\tilde\rho}_{p}
\ee

It is obvious from the equation above that in order to have $\dot{C}=0$  
and $\ddot{C}=0$ we need to have $m \geq 3$. The remaining equations for $B$ are 

\begin{subequations}
\bea
m(m-1){\dot{B}}^2&=&2 \rho_{p} + 2{\tilde\rho}_{p} \\
\ddot{B}+m{\dot{B}}^2 &=& \frac{1+p}{d-2} \rho_{p}
\eea 
\end{subequations}

Assuming a power law ansatz of the form $B(t)=\beta \ln(t)+B_{o}$. Will yield

\begin{subequations}
\bea
\beta&=&\frac{2}{m} \\
e^{-mB_{o}}&=&2\frac{d-2}{\rho_{p}^{o} m (1+p)}
\eea
\end{subequations}

\noindent{\bf{Dilaton Gravity}}

The case $N=1$ has actually been studied in detail by Arapoglu and Kaya \cite{art4}. Quoting verbatim from
their paper (where they took $a=1$ from the start) they use the following ansatz

\begin{subequations}
\bea
&&\phi\,=\, \phi_1\,\ln (t)\,+\,\phi_0,\\
&&B\,=\, b_1\,\ln (t),\\
&&C\,=\, C_0,
\eea
\end{subequations}

we find that gives using the evolution equations \ref{bieq}, \ref{cieq} and \ref{phieq} we get

\begin{subequations}
\bea
&&b_1=\frac{4}{m+3},\;\;\;\;\;\; \phi_1=\frac{2\,(m-3)}{m+3},\\
&&e^{\phi_0}=\frac{4\,(p+2)\,p}{T_w\,(p+1)(m+3)^2},\\
&&e^{(p+1)C_0}=\frac{(p+2)\,T_m}{p\,T_w}
\eea
\end{subequations}

Where $T_{w}=\rho_{o}$ and $T_{m}=\tilde{\rho}_{o}$. With these  the constraint equation \ref{refk1} is identically satisfied. As an be checked the values for $\phi_{1}$ and $b_{1}$ are in accord with $\beta$ in \ref{bphib} and $\varphi$ in \ref{varphib} for $a=1$ meaning $\tau=-1/2$.

Now it is clear that to have a decresing dilaton so as to not enter the strong coupling regime of string
theory in the far future one has to have $m\leq 3$.

\end{document}